\documentclass[twocolumn,iop,apj]{aastex63}
\usepackage{amsfonts,amsmath,graphicx,natbib,url,hyperref,nicefrac}
\usepackage{amssymb, verbatim, subfigure, paralist, soul, comment}

\hypersetup{colorlinks=true, urlcolor=blue, citecolor=blue}

\usepackage{color}
\newcommand{\sprout}{\texttt{Sprout}}
\newcommand{\add}[1]{\textcolor{black}{#1}}

\shorttitle{Anisotropies in SNR} 
\shortauthors{Mandal et al.}

\begin{document}

\title{A 3D Numerical Study of Anisotropies in Supernova Remnants}

\author[0000-0001-9484-1262]{Soham Mandal}
\affiliation{Department of Physics and Astronomy, Purdue University, 525 Northwestern Avenue, West Lafayette, IN 47907, USA}

\author[0000-0001-7626-9629]{Paul C. Duffell}
\affiliation{Department of Physics and Astronomy, Purdue University, 525 Northwestern Avenue, West Lafayette, IN 47907, USA}

\author[0000-0002-1633-6495]{Abigail Polin}
\affiliation{The Observatories of the Carnegie Institution for Science, 813 Santa Barbara St., Pasadena, CA 91101, USA}
\affiliation{TAPIR, Walter Burke Institute for Theoretical Physics, 350-17, Caltech, Pasadena, CA 91125, USA}

\author[0000-0002-0763-3885]{Dan Milisavljevic}
\affiliation{Department of Physics and Astronomy, Purdue University, 525 Northwestern Avenue, West Lafayette, IN 47907, USA}
\affiliation{Integrative Data Science Initiative, Purdue University, West Lafayette, IN 47907, USA}

\email{mandal0@purdue.edu}

\begin{abstract}

We develop a suite of 3D hydrodynamic models of supernova remnants (SNRs) expanding against the circumstellar medium (CSM). We study the Rayleigh-Taylor Instability (RTI) forming at the expansion interface by calculating an angular power spectrum for each of these models. The power spectra of young SNRs is seen to exhibit a dominant angular mode, which is a diagnostic of their ejecta density profile as found by previous studies. The steep scaling of power at smaller modes and the time evolution of the spectra is indicative of absence of a turbulent cascade. Instead, as the time evolution of the spectra suggests, they may be governed by an angular mode dependent net growth rate. We also study the impact of anisotropies in the ejecta and in the CSM on the power spectra of velocity and density.  We confirm that perturbations in the density field (whether imposed on the ejecta or the CSM) do not influence the anisotropy of the remnant significantly unless they have a very large amplitude and form large-scale coherent structures.  In any case, these clumps can only affect structures on large angular scales.  The power spectra on small angular scales is completely independent of the initial clumpiness and only governed by the growth and saturation of the Rayleigh-Taylor instability.

\end{abstract}


\section{Introduction} \label{sec:intro}

Supernova remnants (SNRs) are formed from stellar ejecta driven out by supernova explosions. High-resolution observations have made it possible to study the three-dimensional structure in many SNRs, e.g., Tycho's SNR \citep{Williams+2017ApJ}, N132D \citep{Vogt+2011Ap&SS,Law+2020ApJ}, Cassiopeia A \citep{DeLaney+2010ApJ,Milisavljevic13,MF15}, and SN\,1987A \citep{Larsson16}.

Hydrodynamic \citep{Chevalier+1978ApJ,Velazquez+1998A&A} and magnetohydrodynamic \citep{Jun+1995Ap&SS,Bucciantini+2004A&A} Rayleigh-Taylor Instability (RTI) are expected to be possible factors responsible for these anisotropies. The interaction of the ejecta with the circumstellar material (CSM) drives a forward shock and a reverse shock into the CSM and the ejecta, respectively. The contact discontinuity (CD), or the interface between the shocked ejecta and the shocked CSM has been shown to be Rayleigh-Taylor unstable \citep{Chevalier+1992ApJ}. However, the influence of seed anisotropies on the RTI structures is unclear. Large scale anisotropies have often been attributed to those present in the SN explosion itself, e.g., Tycho's SNR \citep{Ferrand+2019ApJ}, SN 1987A \citep{Orlando+2020A&A}. \add{For instance, the standing accretion shock instability \citep[SASI;][]{Blondin2005} is thought to be one of the primary drivers of large-scale anisotropies in the ejecta structure of core-collapse SNe during the first seconds after the collapse \citep{Blondin+2007Nature,Iwakami+2008ApJ}}. Other important drivers of anisotropies endemic to the SN include jets \citep{GC+2014ApJ} and stochastic processes occurring shortly after core collapse due to events like radioactive decay by $^{56}$Ni plumes and neutrino driven convection \citep{Wongwathanarat+2015A&A,Wongwathanarat+2017ApJ,Orlando+2016ApJ,Orlando+2021A&A,Gabler+2021MNRAS}. In some other cases, it has been argued that dense, asymmetric circumstellar environments are required to explain atypical features in some young SNRs \citep{Celli+2019MNRAS,Sano+2020ApJ_a,Sano+2020ApJ_b}.

\begin{figure*}
\centering
\gridline{\fig{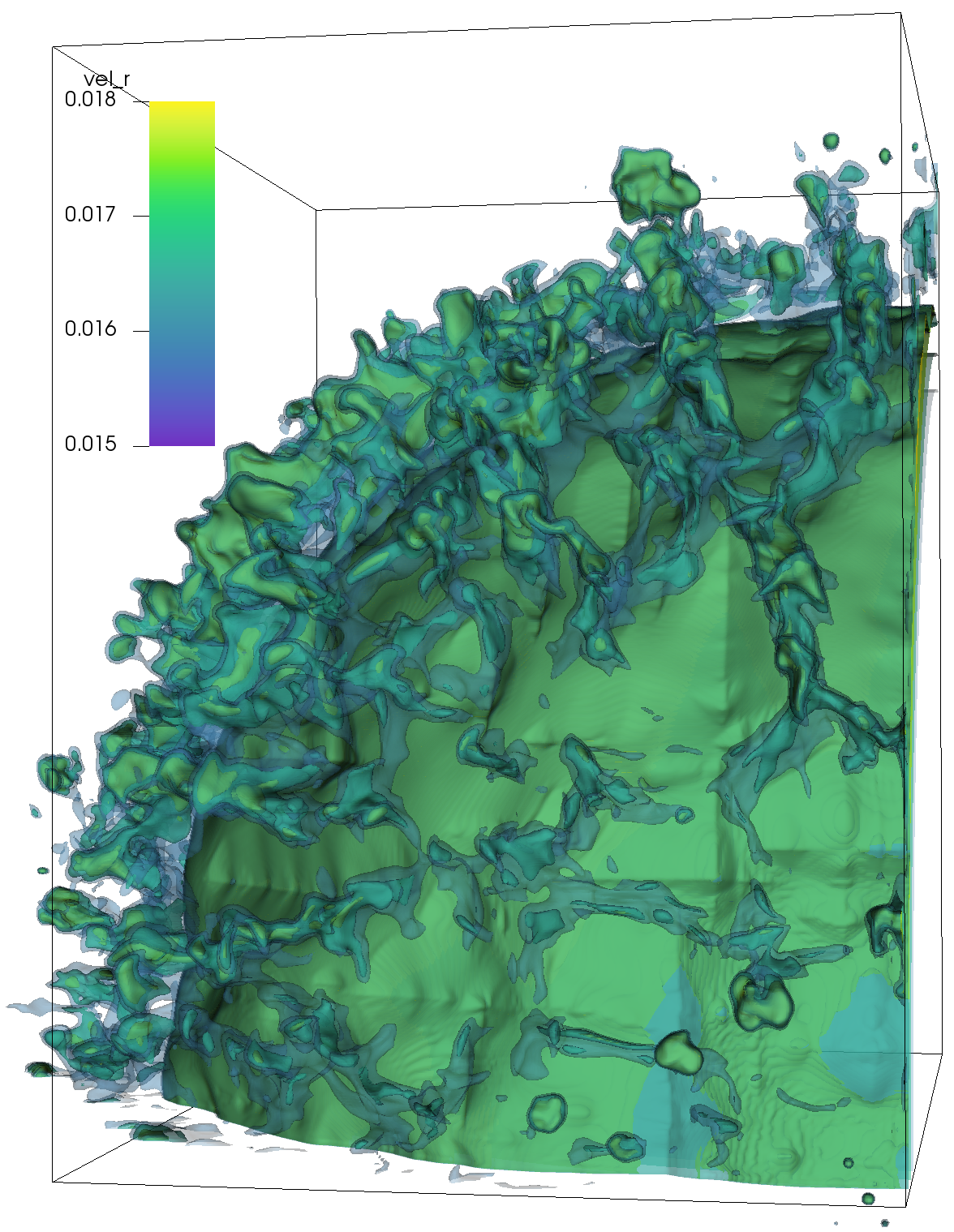}{0.435\textwidth}{}\fig{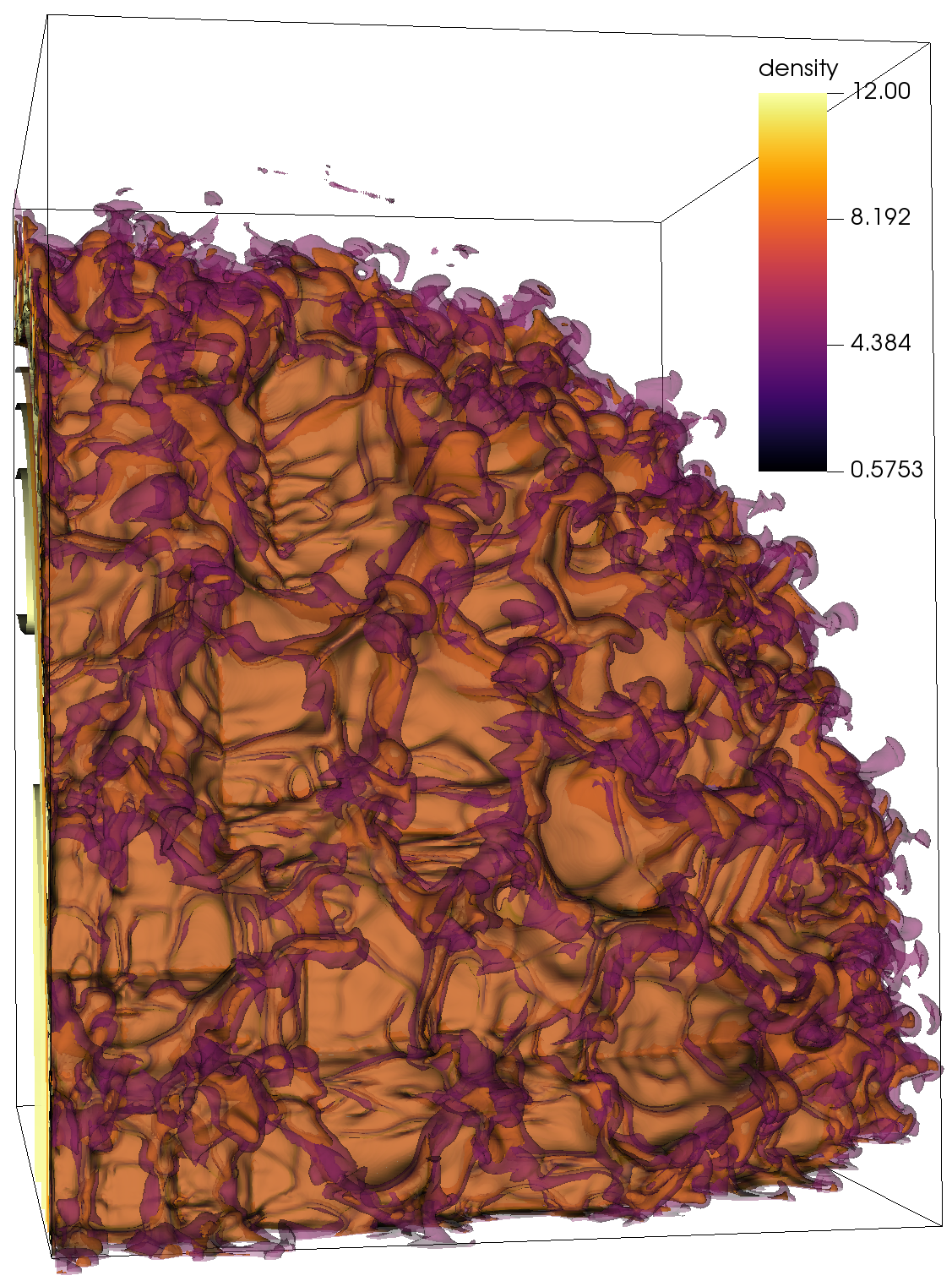}{0.42\textwidth}{}}
\caption{Radial velocity (\textit{left}) and density (\textit{right}) isosurfaces for the $n=9,s=0$ Chevalier self-similar model at a resolution of $(512)^3$. As these represent idealized self-similar solutions, these results apply to any time after the solution has converged to a statistically self-similar state. For a typical Type-Ia SN, they apply to any time after about 30 years and before $10^4$ years, which is when SNR transitions to the Sedov-taylor phase \citep{Woltjer1972}.} The values have been chosen so as to focus on the shocked region between the ejecta and the CSM, which houses the turbulent RTI structures. The visualization has been generated using the open-source software VisIt \citep{VisIt}.
\label{fig:3d_view}
\end{figure*}

\begin{figure*}
\centering
\gridline{\fig{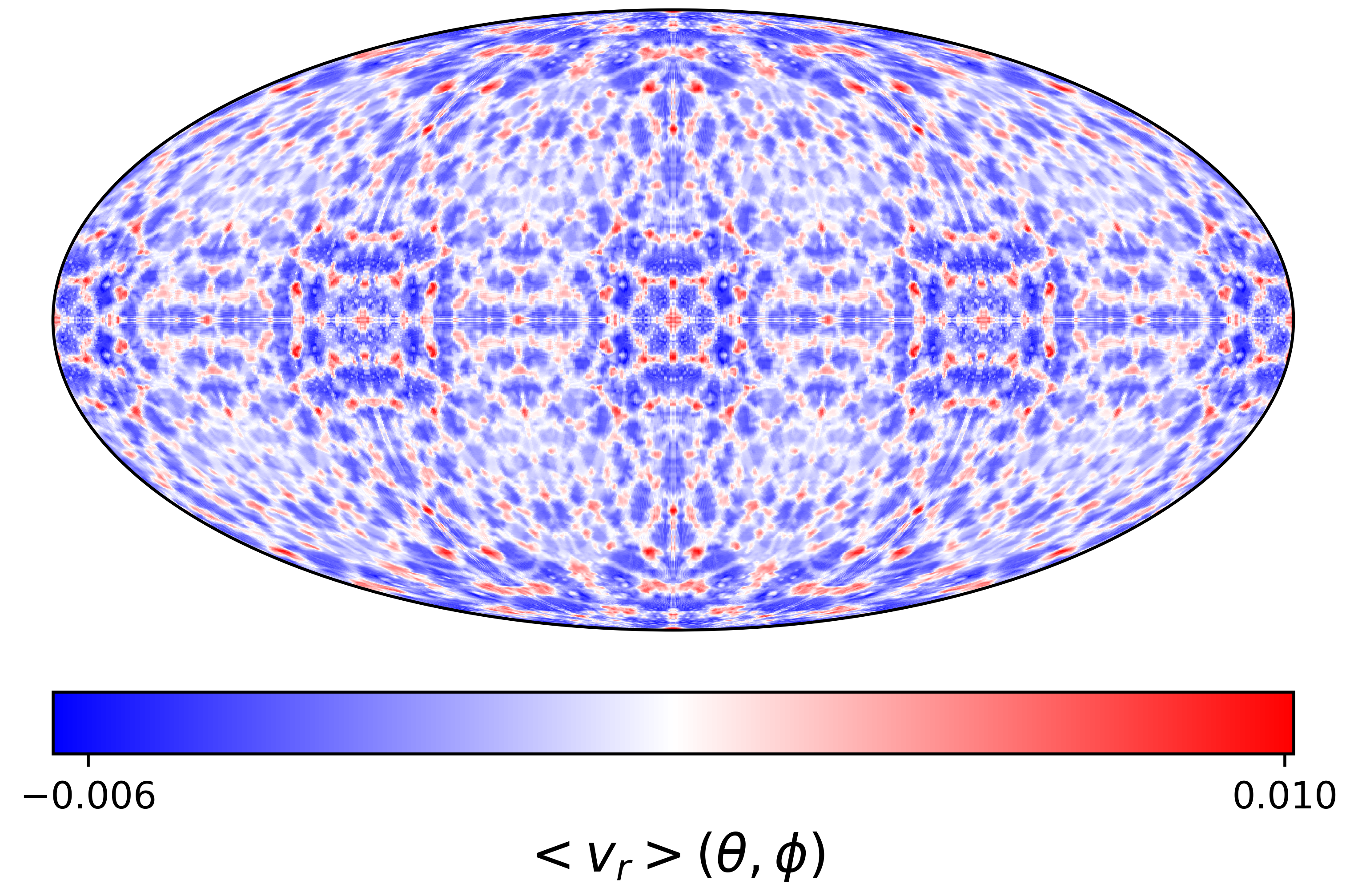}{0.5\textwidth}{}
          \fig{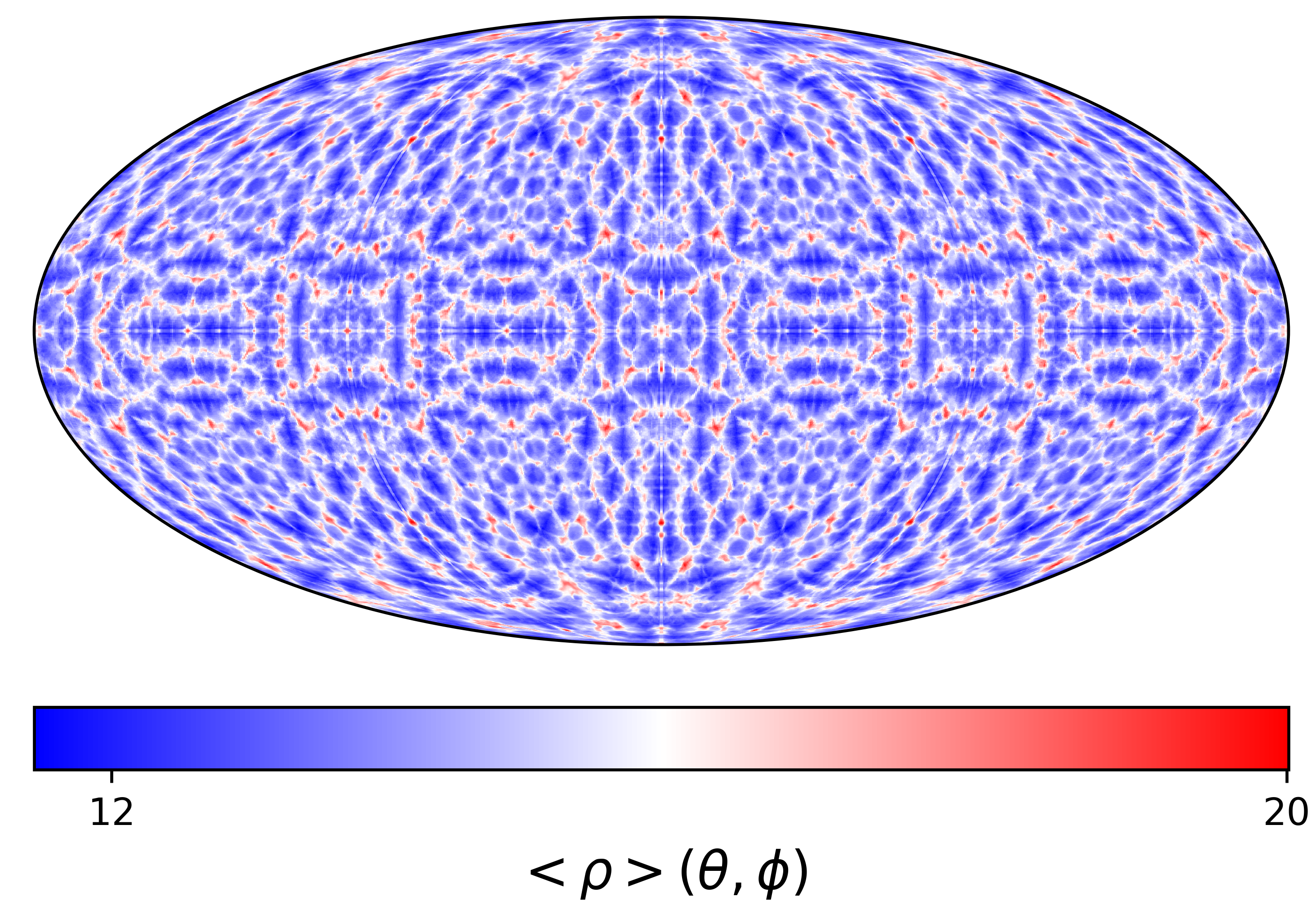}{0.485\textwidth}{}
          }
\caption{Mollweide projections of \textit{(left)} the radial velocity map ($\left<v_r(\theta,\phi)\right>$) and \textit{(right)} the density map ($\left<\rho(\theta,\phi)\right>$) (as defined by Equation \ref{eq:surface_maps}) for the $n=9,s=0$ Chevalier solution. The repeating square-like features (more prominent in the velocity map) are formed by the presence of carbuncle instability in our solutions. They form in directions aligned to the grid and occur due to the fact that the shock isn't aligned with the grid everywhere. Since our solutions are idealized and self-similar, these results apply to any time after the solution has converged to a statistically self-similar state. For a typical Type-Ia SN, they apply to any time between $30-10^4$ years.}
\label{fig:mollweide}
\end{figure*}

Numerical studies of SNRs have identified power spectral analysis as a powerful tool for quantifying anisotropies in the remnant models, e.g., \cite{Warren+2013MNRAS,Ferrand+2019ApJ,Ferrand+2021ApJ,Gabler+2021MNRAS,Polin+2022}. In particular, \cite{Polin+2022} use this technique to identify a dominant angular scale $\theta_0$ in RTI structures formed in SNRs, which separates features formed by turbulence from those endemic to the SN. They find that $\theta_0$ is a function of the density scale height of the ejecta. They also find that the dominant angular scale or the shape of the angular power spectrum of the SNR structure is not affected by fluctuations in the surrounding media in general. \add{However, their value of $\theta_0$ and the shape of the turbulent power spectrum at large wavenumbers does not agree with corresponding results found by \cite{Warren+2013MNRAS}.} In addition, \cite{Polin+2022} obtain their results using 2D axisymmetric numerical calculations, which is in general different from 3D calculations, and doesn't necessarily agree with results from the latter. For example, laboratory experiments and 3D numerical calculations generally exhibit a turbulent cascade of energy from large length scales to smaller ones, a feature that is absent in 2D calculations of turbulence. Hence, we perform a follow-up investigation of hydrodynamic instabilities in SNRs in this work using a suite of 3D numerical models. These models were developed using a new expanding-mesh hydro code called \sprout. In general, this is a complex problem to study  because of the multitude of phenomena involved. SN explosions are often highly asymmetric, as seen for Cas A \citep{Rest+2011ApJ}, W49B \citep{Lopez+2013ApJ,Zhou+2018A&A}, RCW 86 \citep{Broersen+2014MNRAS}, SN1987A \citep{Wang+2002ApJ} and many other SNRs. Moreover, CSM environments may develop large-scale asymmetries and complex density profiles as a result of the mass-loss and evolutionary history of the progenitor system and have to be constrained using theoretical models \citep{Dragulin+2016ApJ} and multi-wavelength observations \citep{Chomiuk+2012ApJ,Cendes+2020ApJ,Sand+2021ApJ}. For instance, there is a growing subclass of SNe that are expected to have run into a circumstellar shell much denser than the ambient medium \citep{Smith+2008ApJ,Jencson+2016MNRAS,Dickinson+2023}, which are formed as a result of a brief period of very high mass loss by the progenitor. Our models do not attempt to simulate all these scenarios explicitly. Instead, we perform a systematic analysis on a range of idealized models to understand the physical laws governing hydrodynamic instability resulting from the expansion of SNR ejecta, and its dependence on system parameters such as the progenitor density structure. We also incorporate seed anisotropies or ``clumps" in the CSM and in the SN ejecta for some of our models via a parameterized model. This allows us to study the effects of these clumps on the turbulent structures in terms of their size and mass. Our models, on account of being mostly spherical, are more directly comparable to remnants of Type Ia SNe since most of them are expected to result from nearly spherical explosions \citep[][and references therein]{Soker2019}. Nevertheless, the conclusions we obtain about the nature of turbulence in SNRs are expected to be general and hold for core-collapse SNRs as well.

This paper is organized as follows. In Section \ref{sec:numerical}, we describe the numerical method employed by \sprout\, briefly along with the model setup. Section \ref{sec:power_spectra} describes the power spectral analysis technique used to study the models. The results are presented in Section \ref{sec:results}. Sections \ref{sec:int_clumps} and \ref{sec:ext_clumps} describe how the SNR structure is affected by the presence of anisotropies in the ejecta and the CSM, respectively. In Section \ref{sec:discussion}, we summarize our results and discuss our expectations for structure of observed SNRs.

\section{Numerical setup} \label{sec:numerical}

\subsection{Numerical method} \label{subsec:method}

\sprout\, \citep{Sprout} solves Euler's equations of ideal fluid dynamics in conservative form on a Cartesian mesh using a second-order upwind Godunov method. Second-order spatial accuracy is obtained through piecewise linear reconstruction of the primitive variables, using a minmod slope limiter. \sprout\, utilizes the moving mesh methodology \citep{Springel2010MNRAS,Duffell+2011ApJS} to expand its mesh with time, preserving all aspect ratios. The mesh motion thus results in all position vectors $\mathbf{r}$ being updated to $\mathbf{r+\delta r}$ after a timestep $\Delta t$ as follows:

\begin{equation}
\label{eq:mesh_motion}
    \mathbf{r+\delta r} = \mathbf{r} + H(t) \left( \mathbf{r}-\mathbf{r_0} \right) \Delta t
\end{equation}

Thus we can resolve fluid motion for several orders of magnitude in distance, although it is only useful to resolve one or two orders of magnitude at any given instant for a reasonable duration of computation time. This also gives us the advantage that each order of magnitude in size should take roughly the same amount of time to solve for, provided the fluid velocity remains constant. 

\begin{figure*}
\centering
\gridline{\fig{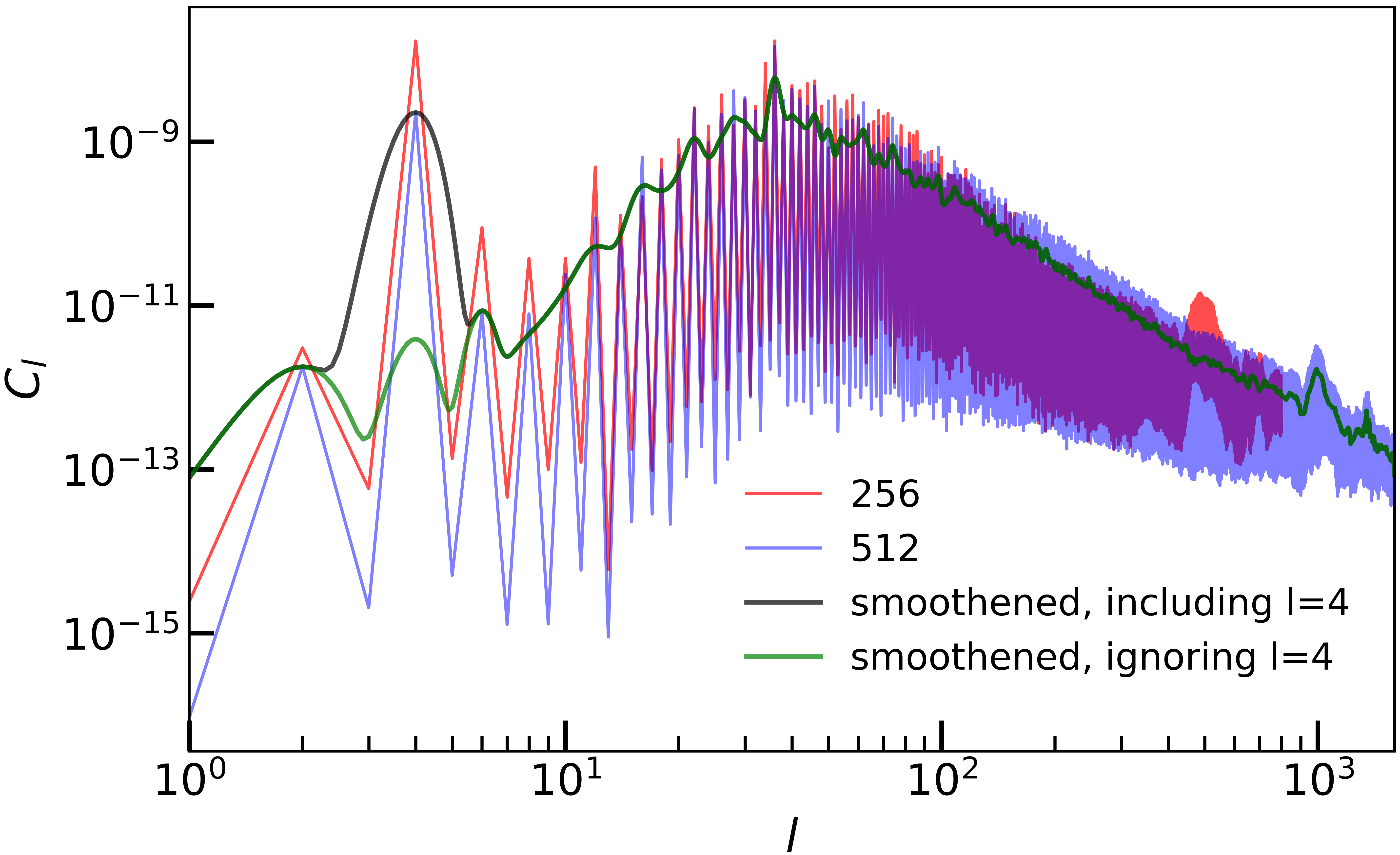}{0.48\textwidth}{}
          \fig{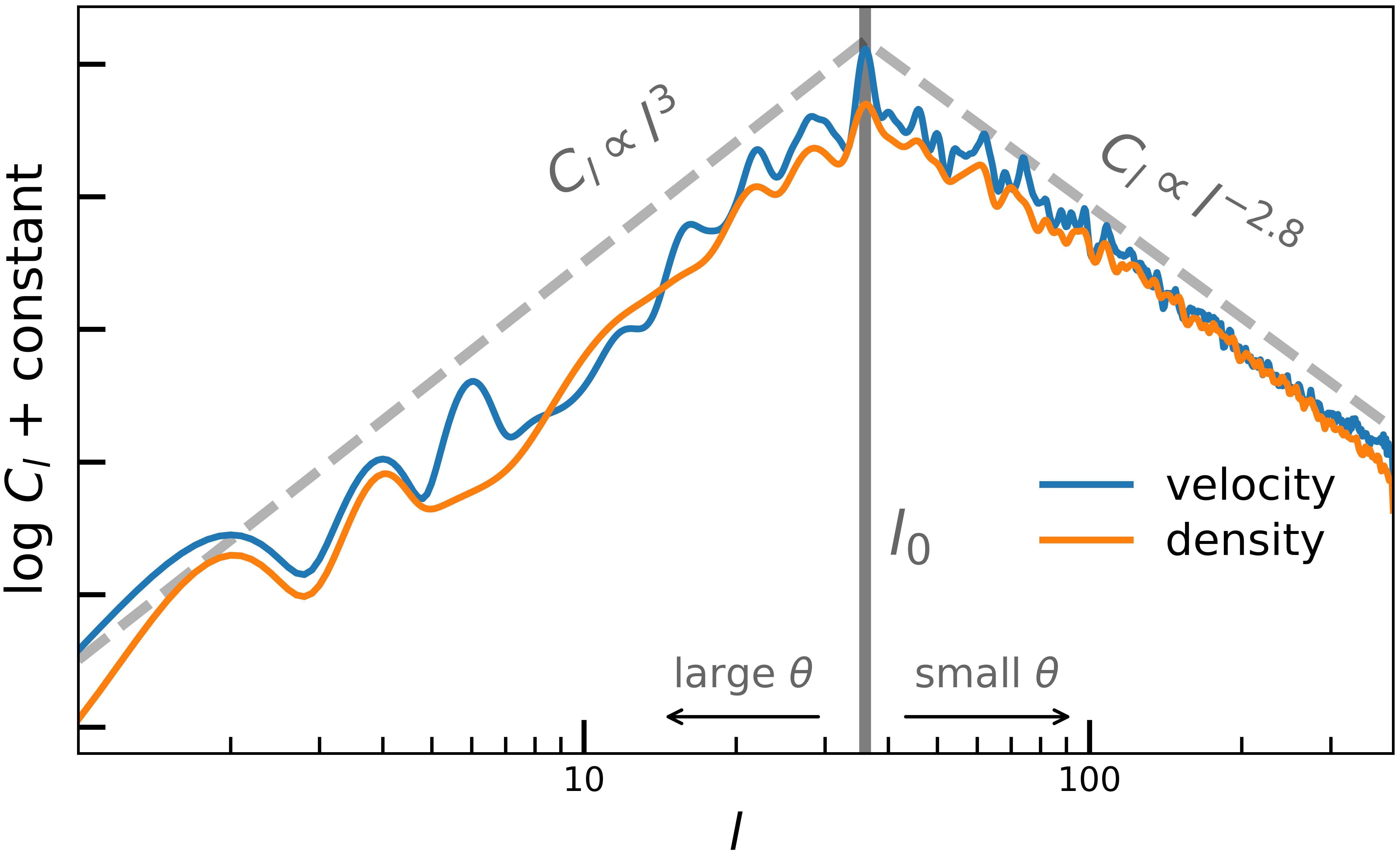}{0.48\textwidth}{}
          }
\caption{Power spectra for the $n=9,s=0$ solution. \textit{(Left)} Velocity power spectra at two different computational resolutions. The peaks at $l=4$ are generated by a mild carbuncle instabilty arising in our solutions. A smaller peak can be seen at $l>100$ in the left panel, caused by the grid scale. It occurs at different locations in the plot for different resolutions. The black line shows a smoothened (by convolving with a Gaussian kernel) version of the spectrum for the $512^3$ run. The green line performs the same operation, but ignores the numerical artifact at $l=4$. The rest of the work uses this version of the spectra for visualization. \textit{(Right)} The smoothened velocity and density spectra (ignoring the $l=4$ peak). They have the shape of a broken power-law, depicted by the grey dashed lines. The vertical line is at the wavenumber where the power-law is broken, denoted by $l_0$. Since our solutions are idealized and self-similar, these results apply to any time after the solution has converged to a statistically self-similar state. For a typical Type-Ia SN, they apply to any time between $30-10^4$ years.}
\label{fig:spectra_intro}
\end{figure*}

\subsection{Initial conditions} \label{subsec:IC}

The outer ejecta in young SNRs are expected to be cold and expanding homologously into a stationary CSM, with a density profile declining steeply with radius. We choose a power-law density profile \citep{Arnett1988ApJ}, in accordance with calculations of adiabatic explosion of polytropes \citep{Matzner+1999ApJ,Chomiuk+2016ApJ}, although there are other models for SN ejecta; for example \cite{Warren+2013MNRAS} employed an exponential ejecta profile consistent with the popular ``W7" model \citep{Nomoto+1984ApJ} in their initial conditions. Even more realistic initial conditions were employed by \cite{Ferrand+2019ApJ,Ferrand+2021ApJ} and \cite{Orlando+2021A&A}, obtained from 3D models of SN explosion. We favor a simple analytical model even though it's not ideal for directly comparing to data, because it allows us to systematically vary the relevant parameters and obtain predictions for physical laws governing their impact on turbulent phenomena in SNRs.

One advantage of using the power-law density profile is that it provides a self-similar solution for the SNR \citep{Chevalier1982ApJ}, allowing the turbulence to reach a statistical steady state (as will be shown in Section \ref{subsec:ev}). After steady state has been attained, the SNR model can be studied at any single instant of time to completely characterize the turbulence, applicable to all times. Our initial conditions are then given by:

\begin{equation}
   \begin{split}
       \mathrm{Ejecta:}\;\;\rho &\propto r^{-n}t^{n-3},\;\mathbf{v}=\mathbf{r}/t \\
       \mathrm{CSM:}\;\;\rho &\propto r^{-s},\;\mathbf{v}=0 \\
       P = 10^{-6}&\rho\;\mathrm{(both\;cases)}
   \end{split}
\end{equation}

We use the above initial conditions to obtain 3D numerical self-similar solutions for supernova remnants. The value of $n$ is expected to be between $7-12$ \citep{Colgate+1969ApJ,Chevalier1982ApJ_b}, and $s$ is either chosen to be $0$ (for a constant density CSM) or $2$ (for a stellar-wind environment). We vary n between $9-11$ for both values of $s$. We choose an adiabatic index $\gamma=5/3$ for all our calculations. The calculations start at some time $t_i$ (in code units), and are continued for four decades, that is, till $t=10^4t_i$. Our calculations assume a fiducial value of $t_i=1\;$day in physical units. Thus, the calculations end at a physical time corresponding to about 30 years. As will be shown in Section \ref{subsec:ev}, we find that the turbulence reaches a steady-state by this time. The solution should remain statistically self-similar after this point, until the reverse shock reaches the core and the mass swept up by the shock becomes comparable to the mass of the ejecta. This is when the SNR transitions from the ejecta-dominated phase to the Sedov-Taylor phase, which typically happens around $10^4$ years after the SN explosion \citep{Woltjer1972}.

The expansion of the computational domain ensures that the position of the contact discontinuity is roughly constant with respect to the domain. This requires

\begin{equation}
    H(t) = \frac{\dot{R_c}(t)}{R_c(t)} = \frac{n-3}{t(n-s)}
\end{equation}

where H(t) is as described by Equation \ref{eq:mesh_motion} and $R_c(t) \propto t^{(n-3)/(n-s)}$ is the speed of the contact discontinuity.

Our runs have a fiducial resolution of $(512)^3$. Most of our computations model only one octant of the spherical SNR to save computation time. For one of the cases (n=9 and s=0), we model the full spherical remnant at the same resolution to ensure we get the same results.

\section{Power Spectral Analysis} \label{sec:power_spectra}

The anisotropies seen in the shocked region in our solutions are best quantified by defining some quantity on the (nearly) spherical ejecta surface. The relative strength in the anisotropies of all angular scales can be obtained by expanding this quantity (which we now call $f(\theta,\phi)$) as a linear combination of spherical harmonics:

\begin{equation}
    f(\theta,\phi) = \sum\limits_{l,m} a_{lm} Y_{lm}(\theta,\phi)
\end{equation}

where $a_{lm}$, or the amplitude of a given harmonic mode can be calculated using:

\begin{equation}
   a_{lm} = \int f(\theta,\phi) Y^{*}_{lm}(\theta,\phi) d\Omega
\end{equation}

There are many possible choices for $f(\theta,\phi)$, including the contact discontinuity and forward shock radii distributions of the remnant \citep{Ferrand+2019ApJ,Ferrand+2021ApJ}. We follow the choice made by \cite{Polin+2022} and obtain spherical surface maps for radial velocity and density integrated along some radial lines of sight originating from the center of the supernovae. They are defined as follows:

\begin{equation}
    \begin{split}
        \left<v_r(\theta,\phi)\right> = \frac{\int P(r,\theta,\phi)\mathbf{v}(r,\theta,\phi)\cdot \mathbf{dr}}{\int P(r,\theta,\phi)dr} \\
        \left<\rho(\theta,\phi)\right> = \frac{\int \rho(r,\theta,\phi) P(r,\theta,\phi)dr}{\int P(r,\theta,\phi)dr} \\
    \end{split}
\label{eq:surface_maps}
\end{equation}

The pressure weighting seeks out the relevant quantities in the shocked region since the fluid is cold everywhere else. The surface maps are then normalized through division by the angle-averaged value (of velocity or density). Given the amplitudes of the harmonics, it is then possible to calculate a power spectrum as a function of the harmonic number $l$:

\begin{equation}
    C_l = \frac{1}{2l+1}\sum_{m=-l}^{+l}\left|a_{lm}\right|^2
\end{equation}

We obtain power spectra for our normalized surface maps using the SHTOOLS package \citep{SHTOOLS}.

\section{Results} \label{sec:results}

Figure \ref{fig:3d_view} shows velocity and density isosurfaces from our 3D solution for $n=9$ and $s=0$. The outermost isosurfaces were chosen to have a value such that they lie around the contact discontinuity. Thus they provide a qualitative view of the typical size of the Rayleigh-Taylor fingers. In Figure \ref{fig:mollweide}, we show the radially integrated velocity and density maps (as defined in Equation \ref{eq:surface_maps}) obtained for this 3D solution. Figure \ref{fig:spectra_intro} shows the resulting power spectra (for the $n=9,s=0$ solution). The left panel shows the velocity spectra obtained at different computational resolutions, demonstrating convergence of the spectra. The spectra are seen to peak at some intermediate spherical harmonic (hereafter $l_0$). There is another peak at $l=4$. This is a numerical artifact arising from a very mild carbuncle instability in our solutions, which is also visible in the Mollweide projections in Figure \ref{fig:mollweide} as small square-shaped structures. They form due to the shock not being aligned to the grid \citep{Quirk1994, Robinet+2000}. We smoothen the spectra by convolving the raw data with a Gaussian kernel (shown in black in the left panel). This generates a peak at $l=4$ with an width approximately equal to that of the smoothing kernel. Since this is a purely numerical feature, we develop another smoothened version of the spectra (shown in green) that ignores the bump at $l=4$. This is useful for identifying features visually and will be used for subsequent plots, although all analysis was done using the raw data. The right panel shows the smoothened velocity and density power spectrum. We find that both power spectra display a broken power law, similar to previous studies \citep{Warren+2013MNRAS, Polin+2022}.  For $l$ smaller than $l_0$, or larger angular scales, the power spectra can be approximated by a power law with a positive slope. For smaller angular scales ($l>l_0$), the spectra behave like a power law with a negative slope.

\subsection{General features of the power spectra}

\begin{figure}
\centering
\includegraphics[width=0.48\textwidth]{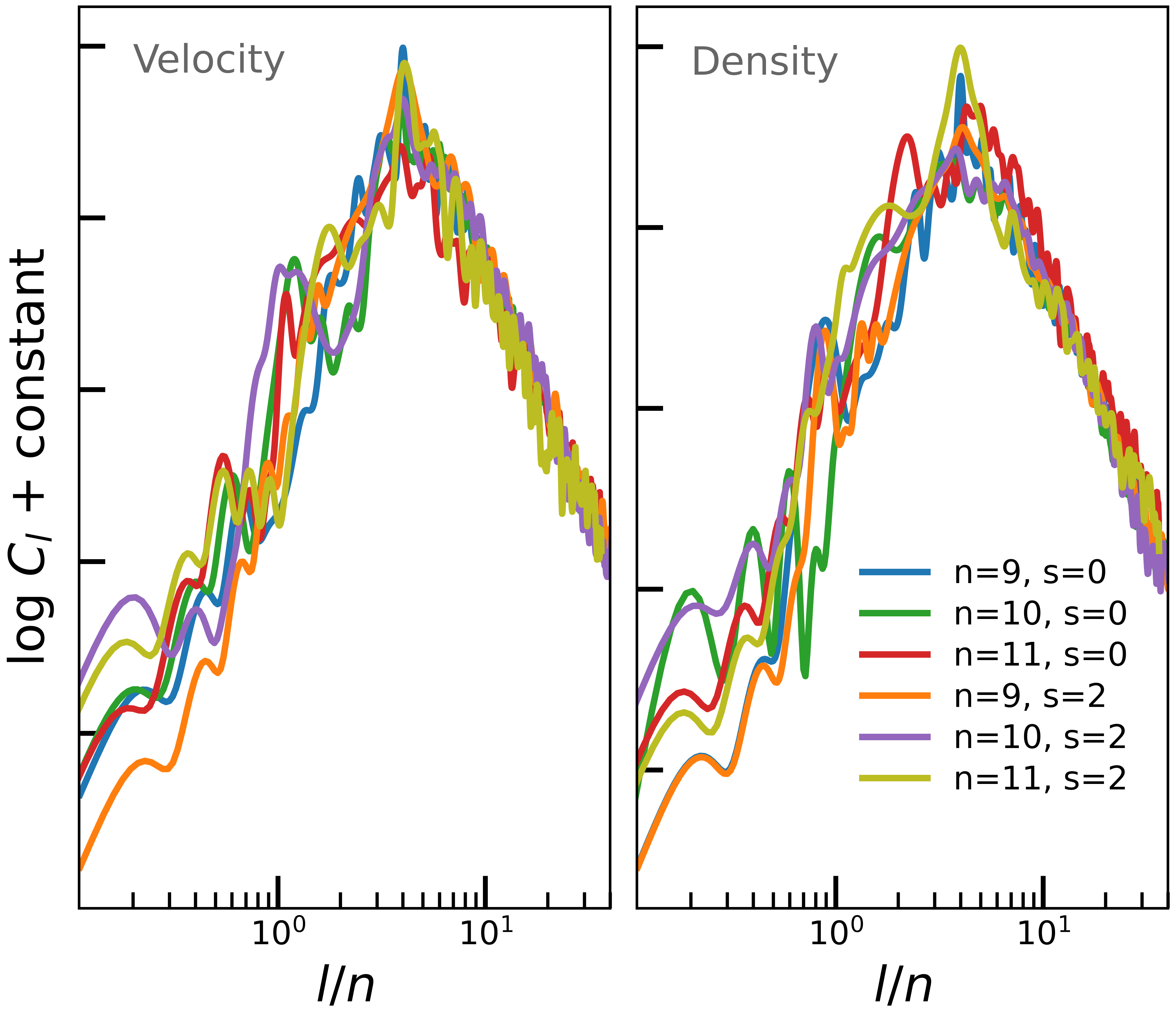}
\caption{\textit{(Left)} Velocity and \textit{(Right)} density power spectrum for all Chevalier models, plotted versus $l/n$. All curves peak at $l/n\sim4$, demonstrating the peak wavenumber $l_0$ is proportional to $n$. As these represent idealized self-similar solutions, these results apply to any time after the solution has converged to a statistically self-similar state. For a typical Type-Ia SN, they apply to any time between $30-10^4$ years.}
\label{fig:all_spectra}
\end{figure}

The velocity and density power spectra for all the different ejecta and CSM density profiles implemented in our models are shown in Figure \ref{fig:all_spectra}. We find that they are fairly consistent with those from previous studies. The break wavenumber, $l_0$ is found to be proportional to the ejecta density power-law index, $n$, same as found by \cite{Polin+2022} and \cite{Warren+2013MNRAS}. The exact value is found to be

\begin{equation}
    l_0 \approx 4n,
\end{equation}

which is close to $\sim3n$ as found by \cite{Polin+2022}. This harmonic corresponds to a dominant angular scale $\theta_0$ ($l_0 \sim \pi/\theta_0$). Our result is consistent with their interpretation that the density scale height of the ejecta,

\begin{equation}
\label{eq:n_def}
    h \equiv \frac{\rho}{\left|\nabla \rho\right|} = \frac{r}{n},
\end{equation}

is roughly of the order of the arc-length $s$ enclosed by the dominant angular scale,

\begin{equation}
    s = r\theta_0 \approx \frac{r\pi}{l_0} \approx \frac{\pi}{4} \frac{r}{n} = \frac{\pi}{4} h.
\end{equation}

\begin{figure}
\centering
\includegraphics[width=0.45\textwidth]{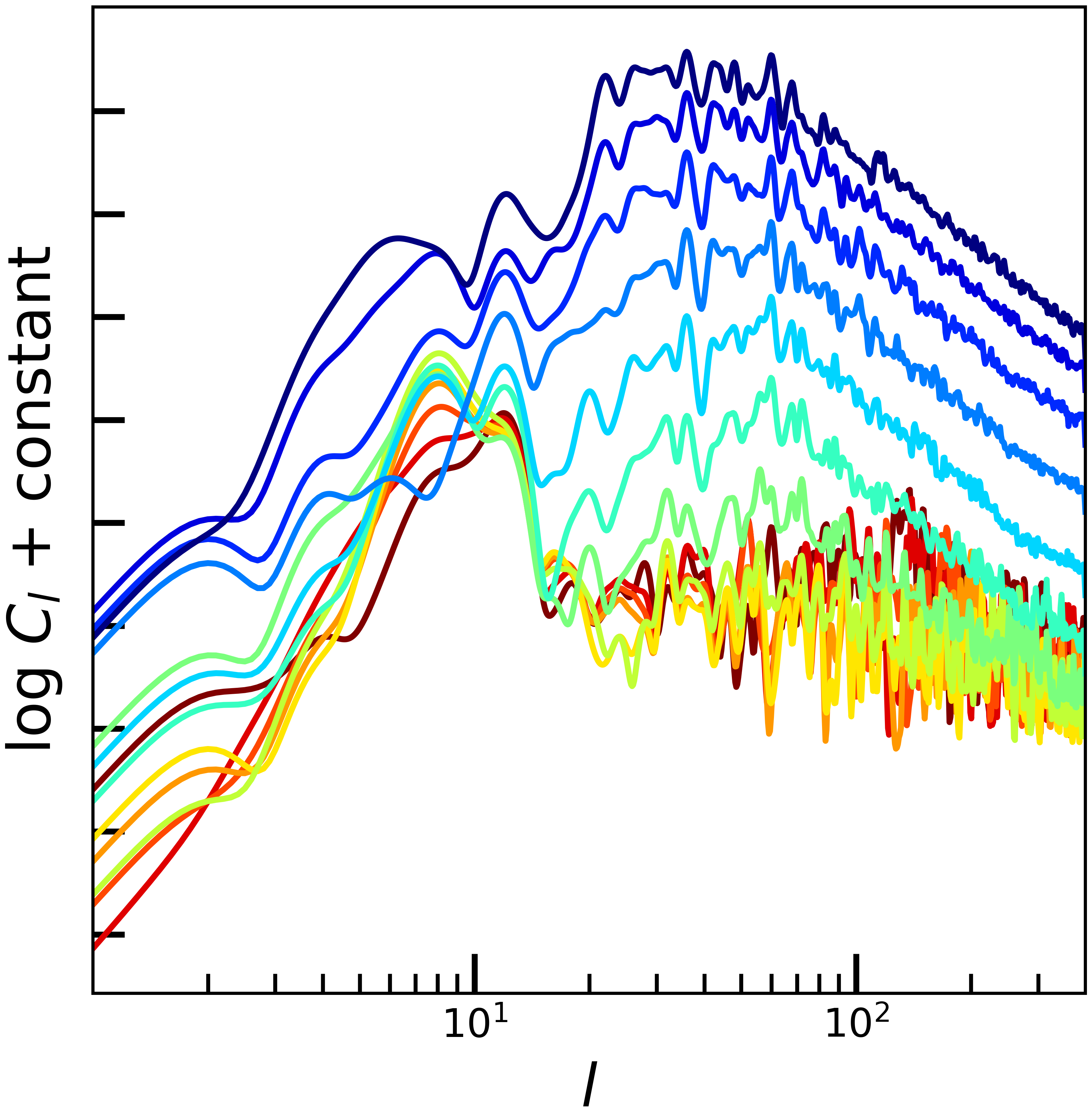}
\caption{Evolution of the velocity power spectrum for the $n=9,s=0$ solution. The power spectrum is evaluated for 12 logarithmically spaced intervals between $t=40\mathrm{d}-30\mathrm{yr}$ (or $t/t_i=40-10000$ in code units)}. The red and the blue curve denote the earliest and the latest instances, respectively. An initially flat spectrum evolves into the peaked structure, indicating maximum net growth at the break frequency.
\label{fig:evolution}
\end{figure}

\begin{figure*}
\centering
\includegraphics[width=0.95\textwidth]{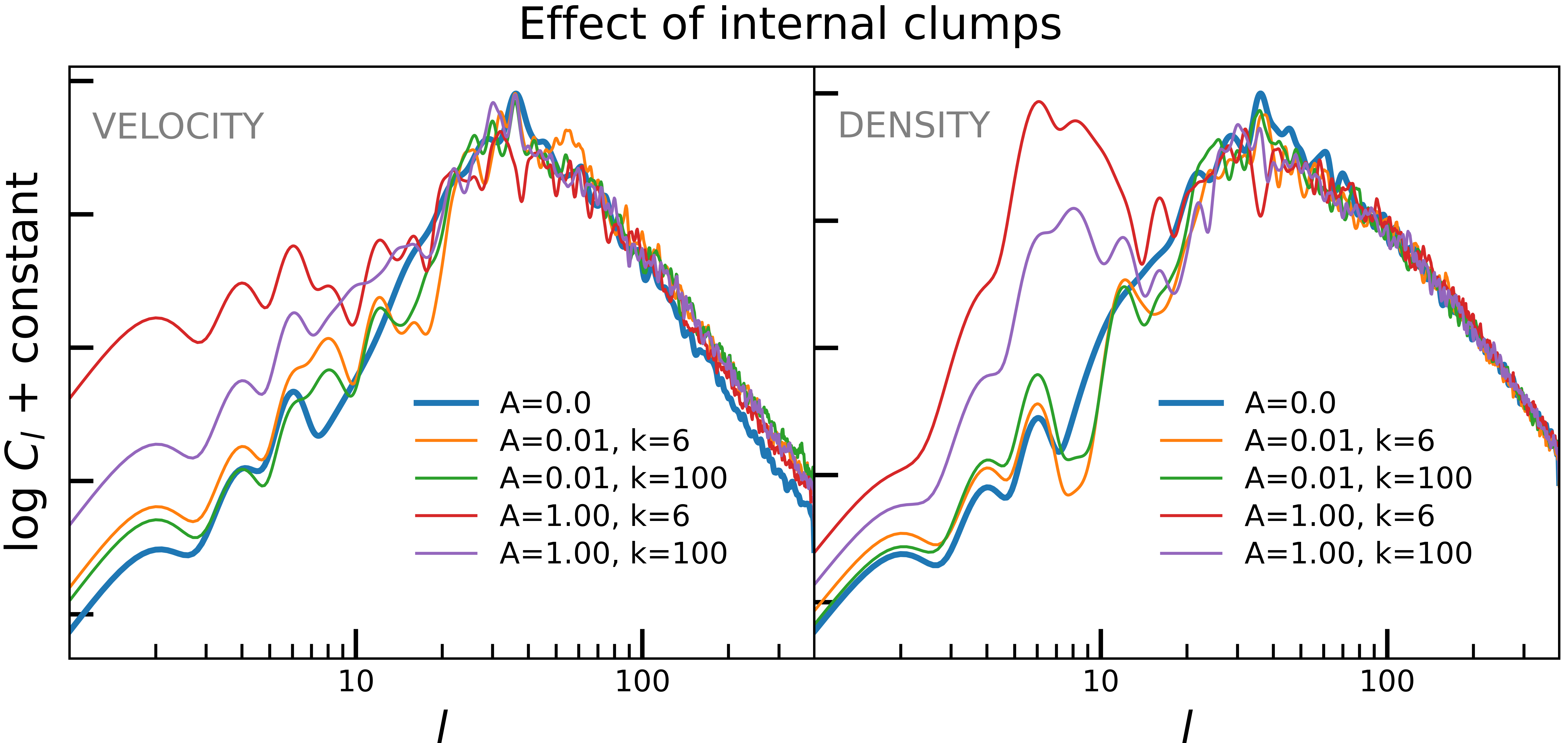}
\caption{Velocity (\textit{left}) and density (\textit{right}) power spectra for models seeded with perturbations in the ejecta according to Equation \ref{eq:perturbations}. The velocity spectra are not heavily impacted by even the strongest of perturbations. The density spectra do show deviation from the fiducial spectrum (for the $n=9,s=0$ model), but only at large angular scales (or small $l$) for perturbations of large amplitude. As these represent idealized self-similar solutions, these results apply to any time after the solution has converged to a statistically self-similar state. For a typical Type-Ia SN, they apply to any time between $30-10^4$ years.}
\label{fig:internal_anisotropy}
\end{figure*}

\cite{Warren+2013MNRAS} also obtain a dominant angular scale in their 3D SNR models with the exponential density profile for the outer ejecta ($\rho \propto e^{-v/v_e}$, $v=r/t$ and $v_e$ being the characteristic velocity of the ejecta). In this case, the density scale height is time-dependent:

\begin{equation*}
    h = v_e t.
\end{equation*}

This leads to an effective value of n (which, from Equation \ref{eq:n_def}, may be interpreted as the ratio between the ejecta radius and the density scale height) that's dependent on time as:

\begin{equation}
    n_{\mathrm{eff}} = \frac{r}{h} = \frac{R_{RS}}{v_e t},
\end{equation}

where the radius of the reverse shock is considered to be the relevant radius.  \cite{Warren+2013MNRAS} obtain results that are consistent with $l_0 \approx 10 n_{\mathrm{eff}}$ (corresponding to a dominant angular scale smaller than we find), as pointed out by \cite{Polin+2022}. This suggests that although the dominant angular scale of SNRs may be dependent on the nature of the density profile of the outer ejecta, such a scale should be expected to arise out of the hydrodynamic interaction between the ejecta and the surrounding CSM. In fact, works starting with 3D initial conditions from thermonuclear SN explosion models \citep{Ferrand+2019ApJ,Ferrand+2021ApJ}, instead of 1D analytical models such as ours, also find a similar break wavenumber or harmonic for their SNR models.

The power spectrum increases as a steep power law ($C_l\propto l^3$) for $l<l_0$ (large angular scales), and then drops off again as a steep power law ($C_l\propto l^{-2.8}$) for $l>l_0$ (small angular scales). The small scale behavior of our models is somewhat different from that found by \cite{Polin+2022} for their 2D axisymmetric models ($C_l\propto l^{-3.5}$), or that obtained by \cite{Warren+2013MNRAS} for 3D models with an exponential density profile ejecta ($C_l\propto l^{-3.9}$). Nevertheless, our power-law scaling is too steep to be consistent with a turbulent cascade \citep{Kolmogorov+1941DoSSR}, in which case the power law should go as $C_l\propto l^{-5/3}$. We note that a power spectrum with scaling of $l^{-3}$ would be consistent with a model where all scales shared the same eddy turnover time. We therefore confirm that RTI structures formed in young SNRs are not likely to exhibit an energy cascade.

\subsection{Time evolution} \label{subsec:ev}

\cite{Polin+2022} put forward the hypothesis that the power spectrum may instead be due to independent growth and saturation of RTI at all scales. We test this hypothesis by looking at the time evolution of the power spectra for a typical SNR model, shown in Figure \ref{fig:evolution}. The spectrum at early times is relatively flat (except for multipole moments described earlier in Section \ref{sec:results}), but gradually peaks at $l_0$ with time, with a steep power law ($C_l\propto l^{-2.8}$) for large $l$ or small scales. Since the power spectrum should scale in general as

\begin{equation}
    C_l \sim |\delta v|^2/l,
\end{equation}

where $\delta v$ is the typical speed of eddies at wavenumber $l$, it follows that for our small scale power law

\begin{equation}
    \begin{split}
        |\delta v| &\sim \frac{\lambda}{\tau} \sim l^{-0.9} \\
        \mathrm{or,\,} \tau &\sim l^{-0.1}, \\
    \end{split}
\end{equation}

where we have used the eddy size $\lambda$ is inversely porportional to $l$ and $\tau$ is the eddy turnover time. Hence, the eddy turnover time is almost the same for eddies of all sizes with $l>l_0$. This is in contrast with Kolmogorov turbulence, where the eddy turnover time scales as $\tau \sim l^{-2/3}$. In the latter case, small scale eddies live longer, allowing energy to be transferred to them from larger scales, where most of the energy initially resides. This is not possible for our results, since all eddies share roughly the same lifetime. Thus, we may not be seeing a turbulent cascade. Instead, each angular mode in our model could be experiencing its own independent net growth rate, dictated by the growth and saturation of RTI at that mode. The dominant mode $l_0$ therefore has the largest net growth rate. We also note that the net growth rate for modes $l<l_0$ is relatively smaller. Fluctuations at these modes persist for much longer than at modes $l>l_0$, where initial perturbations are drowned much faster. We find that the SNR has to expand by $\sim40$ times before developing a steady-state velocity power spectrum. The density spectrum takes much longer to reach steady-state, requiring an expansion by $\sim100$ and $\sim200$ times for the $l>l_0$ and $l<l_0$ modes, respectively. These expansion factors can be compared to a typical young SNR, say Tycho's SNR, which has an age of $\sim450$ years. It is estimated to have a radius of the order of a few parsecs. Assuming a white dwarf progenitor with radius $10^4$ km for the Type Ia SN responsible for Tycho's SNR, we find an expansion factor of $>10^{10}$ for the radius, which should be sufficient for developing steady-state angular power spectra.

\begin{figure*}
\centering
\includegraphics[width=0.95\textwidth]{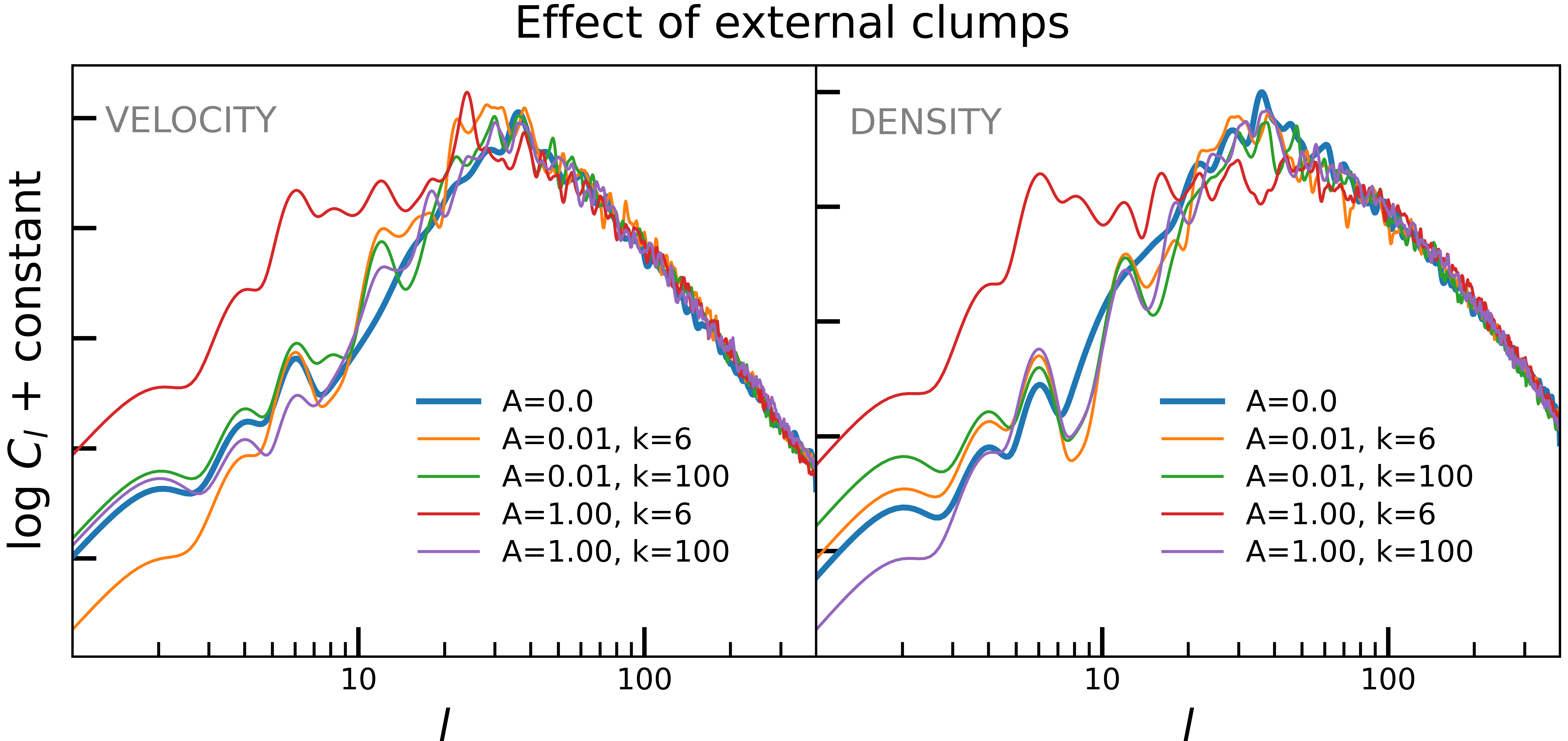}
\caption{Velocity (\textit{left}) and density (\textit{right}) power spectra for models that encounter clumpy CSM, with the clumps defined as in Equation \ref{eq:perturbations}. The velocity spectra aren't influenced by perturbations, as with anisotropic ejecta. The density spectra are only affected by the strongest perturbations. As these represent idealized self-similar solutions, these results apply to any time after the solution has converged to a statistically self-similar state. For a typical Type-Ia SN, they apply to any time between $30-10^4$ years.}
\label{fig:external_anisotropy}
\end{figure*}

\section{Clumpiness in the ejecta} \label{sec:int_clumps}

Numerical studies \citep{Wang+2001ApJ,Orlando+2012ApJ} have shown anisotropies in supernova ejecta can significantly affect the behavior and size of Rayleigh-Taylor fingers and thus leave their imprint on the SNR structure. In fact, observations of clumpy metal distributions in some remnants suggest this is a likely case, e.g., SN 1006 \citep{Winkler+2014ApJ}. In this section, we discuss the results of our investigation on the survival of density perturbations in the supernova ejecta of our models. We add the following density perturbations to the ejecta in our $n=9,\,s=0$ solution:

\begin{equation}
\label{eq:perturbations}
    \delta \rho = \rho_0\, \mathrm{exp}\left[ A\mathrm{sin}(k\;\mathrm{ln}\,r)\,\mathrm{sin}(k\theta)\,\mathrm{sin}(k\phi)\right].
\end{equation}

Figure \ref{fig:internal_anisotropy} shows the effect of varying amplitudes ($A$) for perturbations at a low spatial frequency harmonic ($k=6$) and a high spatial frequency harmonic ($k=100$). The large wavenumber ($l>l_0$) behavior of both the velocity and the density spectra are essentially independent of the nature of fluctuations in the ejecta density profile. The density spectra show deviation from the fiducial ($A=0$) spectrum at small wavenumbers ($l<l_0$), especially for $A=1.0$. This is consistent with our observation of the time evolution of the power spectra, where we find that the small wavenumber modes persist for longer durations. The velocity spectrum on the other hand, is only mildly affected, even for small $l$. It is interesting to note that even for small scale anisotropies ($k=100$), the fluctuations only show up at small wavenumbers. We do not yet have an explanation for this behavior. Overall, our findings would imply that the small-$l$ component of the density spectrum is acutely sensitive to anisotropies in the ejecta.

\begin{figure*}
\gridline{\fig{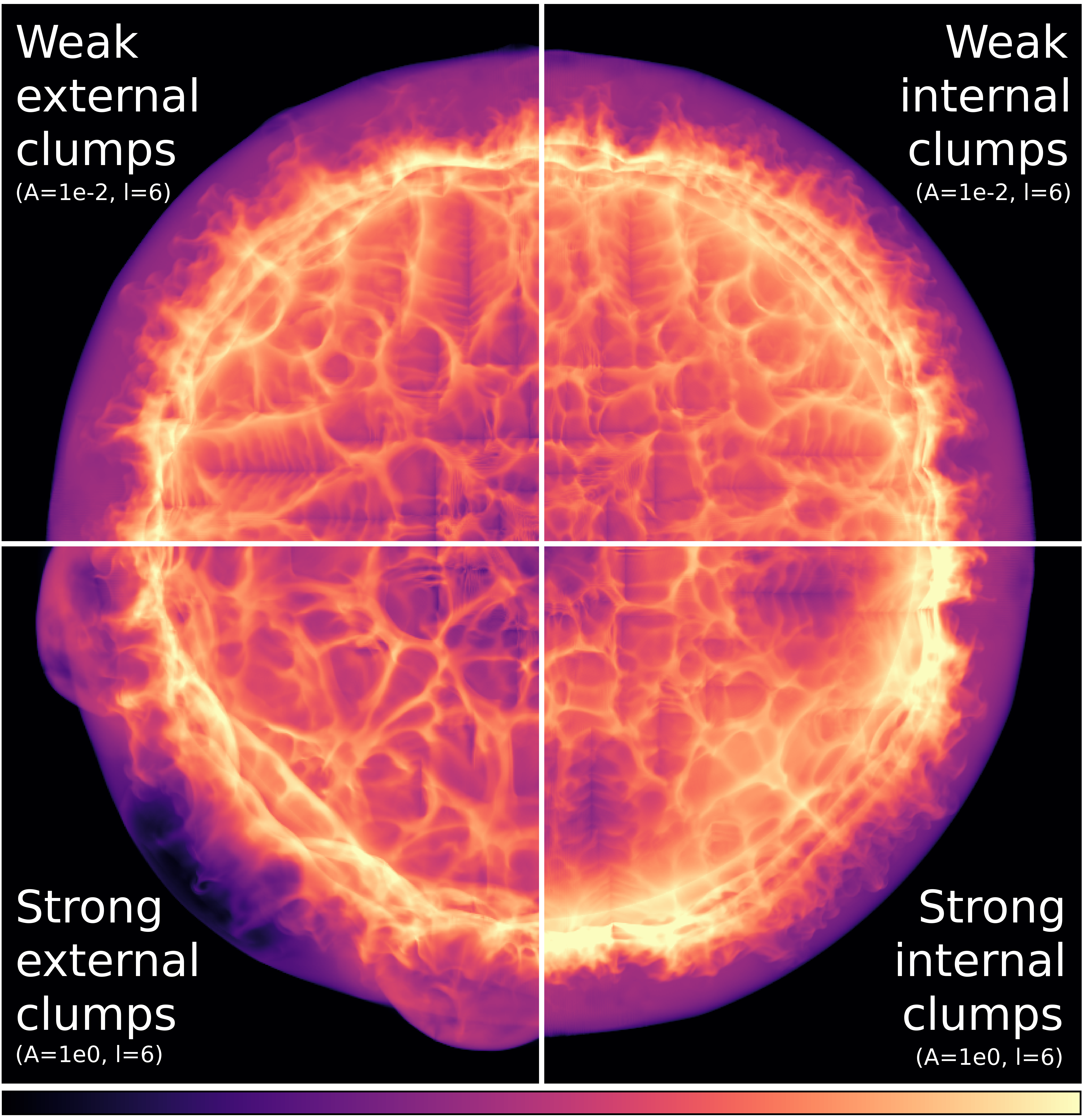}{0.49\textwidth}{}
          \fig{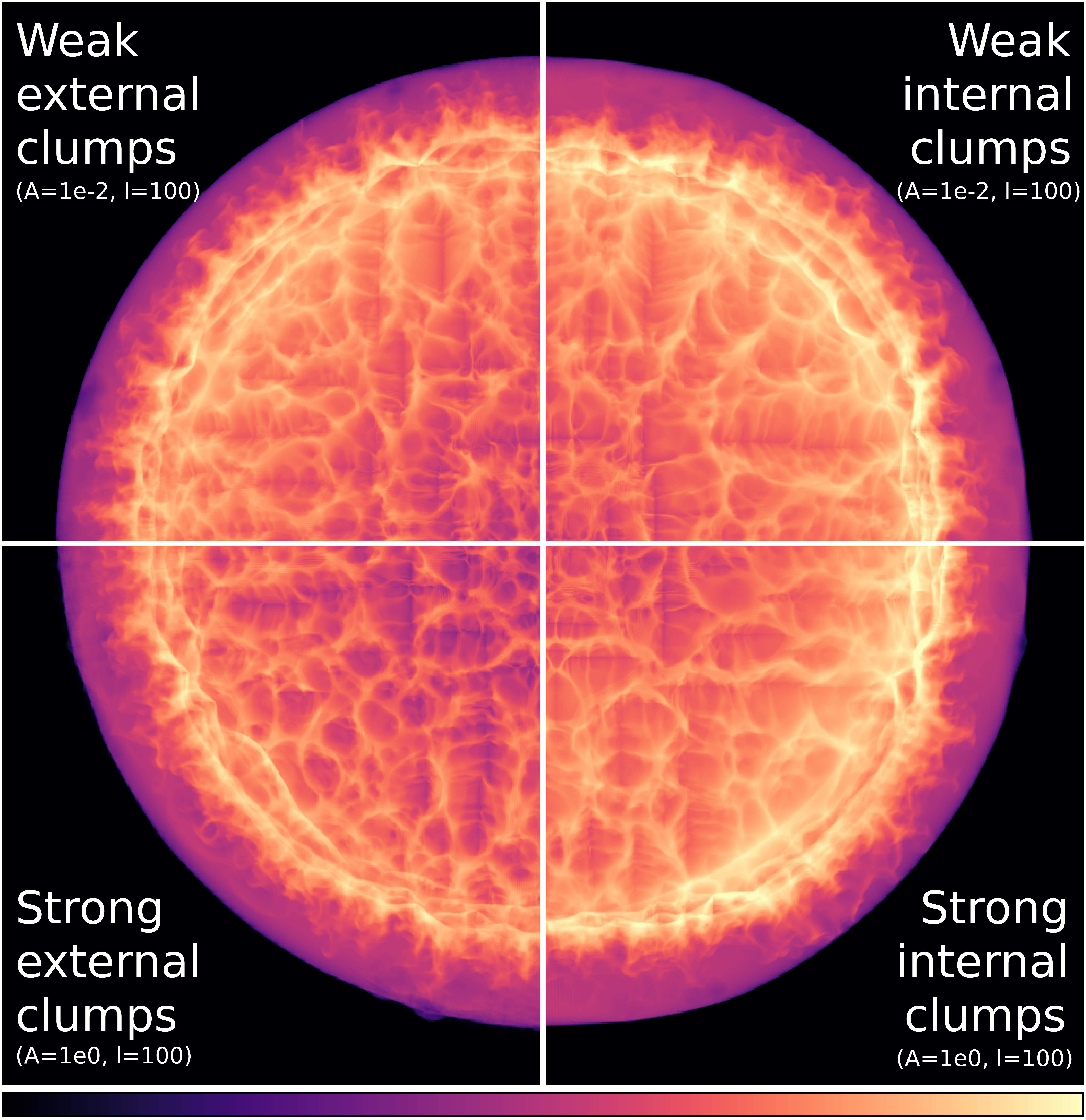}{0.49\textwidth}{}
          }
\centering
\caption{\add{The square of the density of the shocked plasma integrated along the line of sight.} These images were developed using our models with $n=9$, $s=0$, and different initial perturbations to the density field, assuming the plasma is optically thin and is heated to over $10^6K$ in the shocked region. (\textit{Left}) Perturbations for these models are large-scale (corresponding to $l=6$) and are either internal or external to the SN explosion. The quantity $A$ is as defined in (\ref{eq:perturbations}). It can be seen large structures persist in the SNR only if the clumps are massive (bottom panels), in addition to being large. (\textit{Right}) The initial density perturbations for these models are on much smaller length scales ($l=100$). No clumps of the corresponding size are seen in the SNR morphology, irrespective of the mass. Therefore in each case, the dominant angular scale seen is purely a function of the density scale height. As these represent idealized self-similar solutions, these results apply to any time after the solution has converged to a statistically self-similar state. For a typical Type-Ia SN, they apply to any time between $30-10^4$ years.}
\label{fig:clumps}
\end{figure*}

\section{Clumpiness in the CSM}  \label{sec:ext_clumps}

The departure from spherical symmetry for many SNRs have also been attributed to clumpy surrounding medium, as opposed to anisotropies in the explosion itself. In the case of Cas A, \cite{Sato+2018ApJ} and \cite{Orlando+2022A&A} argue that the radially inward velocity in parts of the reverse shock and the offset between forward and reverse shock centers require interaction of the SNR with an asymmetric environment. \cite{Fang+2018MNRAS} propose that anisotropies in the surrounding stellar wind may be able to account for anisotropies seen in Tycho's SNR. Here we systematically investigate the effects of such clumpy CSM on our results. We again introduce density perturbations of the form (\ref{eq:perturbations}), but in the CSM instead of the ejecta, and calculate solutions for $n=9,\,s=0$ as in Section \ref{sec:int_clumps}. The values of $\delta$ and $k$ remain the same as before. The resulting spectra are shown in Figure \ref{fig:external_anisotropy}. We find the the nature of the anisotropies in the CSM do not affect the velocity power spectra irrespective of the magnitude of density fluctuations. The density spectra are also affected very mildly, unlike the case of clumpy ejecta. The only noticeable departure is seen when the density perturbations are of the order of unity, as found previously by \cite{Polin+2022}.

\add{Figure \ref{fig:clumps} shows the square of the density of the shocked plasma from a few different models, integrated along the line of sight. They are examples of a first approximation of thermal emissivity map of SNRs, assuming emission is produced only by the shocked plasma, and the plasma is optically thin.} The left half shows images for models with large-scale initial perturbations (with $l=k=6$) and illustrates the difference between structures caused by large external clumps and large internal clumps. It can be seen the large-scale structures are only formed when the clumps are massive (bottom panels, left half). Also, the forward shock can be perturbed when there are large-scale massive clumps in the CSM. In contrast, no such large scale structures or shock perturbations are seen in the right half of Figure \ref{fig:clumps}, which shows thermal emission from models perturbed with small scale clumps ($l=100$). The dominant angular scale is the same for those models, and is solely determined by the density scale height.

\section{Discussion} \label{sec:discussion}

We apply the technique of power spectral analysis to 3D hydrodynamic numerical models of young supernova remnants to probe the nature of anisotropies they exhibit. We study 
the impact of seed anisotropies or clumps in the CSM for our models to verify the well-studied role of these CSM clumps in generating the turbulent structures in SNRs. In addition, we perform a systematic parameter study of anisotropies endemic to the SN explosion to determine the conditions under which they can impact the SNR structure. We confirm that the power spectra of spatial distribution of either density or radial velocity of young SNRs has the form of a broken power-law, as found previously  \citep{Warren+2013MNRAS, Ferrand+2019ApJ, Polin+2022}. The break (or peak) harmonic $l_0$ represents the angular scale where most of the power resides. In summary, we find the following characteristics for our 3D calculations:

\begin{enumerate}
    \item The dominant angular mode ($l_0$) of the power spectrum is related to the power-law index ($n$) of the ejecta density profile as $l_0\sim4n$. Thus, this is a direct diagnostic of the density profile of the outer layers of the SN ejecta, as pointed out by \cite{Polin+2022}. However, the dominant angular mode we obtain is different from that obtained by \cite{Polin+2022} ($l_0\sim3n$) or \cite{Warren+2013MNRAS} ($l_0\sim10n$). More studies are required to find an agreement on what dominant angular mode is to be expected for observations of SNRs.
    
    \item The power spectrum falls off as $C_l\propto l^{-2.8}$ at large $l$ (or small angular scales). This is somewhat shallower than the scalings found by \cite{Warren+2013MNRAS} and \cite{Polin+2022}. The small scale behavior of the power spectrum is found to be very consistent across all of our models, irrespective of the size, strength, or location of seed perturbations introduced.
    
    \item Consistent with previous findings, the power spectra of our 3D SNR models are minimally affected by perturbations or clumpiness in the surrounding medium. The only impact is seen at small wavenumbers ($l<l_0$) in the density spectrum, as in the case of clumpy ejecta, but for very strong perturbations.

    \item In contrast with the previous point, we find that density power spectra of SNRs are quite sensitive to large scale ($l<l_0$) asymmetries in the ejecta. They are found to impart a long-lasting imprint upon the density power spectrum. Fluctuations at small scales subside quickly. The velocity spectra are even less affected irrespective of the angular mode of fluctuations.
\end{enumerate}

In summary, our study confirms known or suggested properties of the angular power spectrum of SNRs, and establishes their connections to intrinsic properties of the ejecta. Although most of these connections were pointed out by \cite{Polin+2022} in the same form, 2D numerical studies of turbulence aren't necessarily accurate descriptions of turbulence seen in nature, as mentioned earlier. \cite{Warren+2013MNRAS} identify the dominant angular scale in their 3D models as solely a function of a scaled age for a few different times, but without an overall prediction for how this power spectrum should evolve with time generally. Hence, it is difficult to establish a direct connection of the observed dominant angular scale for a given SNR to the ejecta properties based on this model. Our discovery thus provides a path to infer density profiles of ejecta interacting with the reverse shock for observations of young SNRs by constructing their angular power spectra, in addition to the prediction that the presence of low harmonic modes in the power spectra are likely indicators of anisotropies endemic to the SN itself, rather than the CSM.

There are other, arguably more realistic choices for SN ejecta models, ranging from spherically symmetric analytical profiles to 3D profiles obtained from numerical solutions of SN explosions. A preferred choice for modeling Type Ia SN ejecta  \citep{Dwarkasdas2000ApJ,Warren+2013MNRAS} is the exponential density profile \citep{Dwarkadas+1998ApJ}, designed to approximate the W7 carbon deflagration model for thermonuclear SN explosions \citep{Nomoto+1984ApJ}. 3D density profiles, obtained from numerical calculations of thermonuclear SN explosions \citep{Seitenzahl+2013MNRAS,Fink+2014MNRAS,Tanikawa+2018ApJ}, are also used to model SNRs. One of the advantages in this case is that one can incorporate realistic large scale anisotropies in the ejecta as a part of the initial conditions \citep{Ferrand+2019ApJ,Ferrand+2021ApJ}. Another point worth mentioning is that our models do not take into account the cooling down of the shocked region via cosmic ray acceleration. A common way to mimic shock cooling in hydrodynamic calculations is to increase fluid compressibility by reducing the adiabatic index \citep{Blondin+2001ApJ,Warren+2013MNRAS}. \cite{Ferrand+2010A&A} explicitly study the impact of cosmic ray cooling on instabilities in SNRs using hydrodynamic models that include a kinetic prescription for particle acceleration at the shock front. The presence of magnetic fields in the SNR is another important factor that could potentially impact our results. For instance, it has been shown that magnetic tension can suppress Rayleigh-Taylor instability, depending on the field strength and orientation \citep{MacLow+1994ApJ,Jones+1996ApJ,Bucciantini+2004A&A,Fragile+2005ApJ}. On the other hand, \cite{Porth+2014MNRAS.443,Porth+2014MNRAS.438} find magnetic fields have minimal effect on the turbulent structure in their numerical models for pulsar wind nebulae, due to strong magnetic  dissipation, and the inability of magnetic fields to suppress unstable modes that exchange fluid between the field lines \citep{Stone+2007ApJ}. A systematic examination of models that incorporate these features is thus required to understand their impact on our results as well as observations and will be pursued in a future study.

Our findings indicate that large scale features in density profile of young SNRs, being acutely sensitive to perturbations in the ejecta, are better tracers of the inherent anisotropy of the explosion, and therefore are not predictable by Rayleigh-Taylor instability alone. The large scale structures in observed remnants may be indicative of the angular scale of the traces of perturbative activity (such as formation of $^{56}$Ni clumps) in the SN ejecta under consideration. On the other hand, small scale features in the density and velocity field are robust to fluctuations, both in the SN ejecta and in the surrounding medium. \cite{Ferrand+2021ApJ} come to a very similar conclusion. They find that the low $l$ modes in the SNR power spectra are dominated by asymmetries endemic to the SN explosion (used as 3D initial conditions for their SNR models), whereas large $l$ (small angular scale) modes form due to RTI alone. We therefore expect the break wavenumber $l_0$, and the power-law index for large values of $l$ to be largely consistent between SNRs, provided extreme events like jets, highly asymmetric explosions, or collisions with a companion do not overwhelm the small-scale structure. We note that even though $l_0$ may not be the peak wavenumber due to the presence of large-scale features, the power spectrum is nevertheless expected to have a break around $l_0$. This motivates a search for dominant angular scales in the morphologies of observed SNRs. Some work has already been done in this direction. For example, power spectra have been developed from resolved x-ray images of Tycho's SNR using the azimuthal distribution of the Rotation Measure \citep{Lerche+1979A&A} and measures of remnant radius \citep{Warren+2005ApJ}. A different direction was  taken by \cite{Shimoda+2018MNRAS}, who construct correlation functions of the synchrotron emissivity maps of Tycho's SNR. \cite{Lopez+2009ApJL,Lopez+2009ApJ,Lopez+2011ApJ} analyze resolved \textit{Chandra} images of a sample of nearby SNRs using multiple techniques, including a 2D multipole expansion, two-point correlation analysis, and wavelet transforms. They demonstrate that core collapse SNe exhibit more large-scale asymmetry (stronger low $l$ modes) than Type Ia SNe in general. It's perhaps worthwhile to mention here that an approach similar to the wavelet transform, called the $\Delta$-variance method, was developed by \cite{Arevalo+2012MNRAS} to extract power spectra for 2D images as well as 3D data sets. However, most of these works start with 2D images, which inherently suffer from projection effects and therefore may not be able to provide accurate three-dimensional information. This problem may be circumvented by deprojecting observations to obtain estimates of 3D density and velocity distribution of the SN ejecta, or by using high resolution imaging and spectroscopy to directly estimate velocities \citep{DeLaney+2010ApJ,Williams+2017ApJ,Seitenzahl+2019PRL}. Therefore, more work on observations and analysis techniques are needed to learn about the underlying physical properties of SN explosions and progenitors by comparing power spectra of observations to that of theoretical models.

\begin{acknowledgments}

We thank Carles Badenes, Donald C. Warren, Vikram Dwarkadas, and Nirupam Roy for helpful comments. We also thank the anonymous referee for their careful reading and constructive criticism. Numerical calculations were performed in part on the Stampede2 supercomputer under allocations TG-PHY210027 \& TG-PHY210035 provided by the Advanced Cyberinfrastructure Coordination Ecosystem: Services \& Support (ACCESS) program, which is supported by National Science Foundation grants \#2138259, \#2138286, \#2138307, \#2137603, and \#2138296.  Calculations were also carried out using the Petunia computing cluster hosted by the Department of Physics and Astronomy at Purdue University.  D.~M.\ acknowledges NSF support from grants PHY-1914448, PHY-2209451, AST-2037297, and AST-2206532.

\software{\sprout\, \citep{Sprout},  
    VisIt \citep{HPV:VisIt}, 
    SHTOOLS \citep{SHTOOLS},
    Matplotlib \citep{matplotlib}.
}

\end{acknowledgments}

\appendix

\section{Numerical convergence studies}

\begin{figure}
\centering
\includegraphics[width=0.45\textwidth]{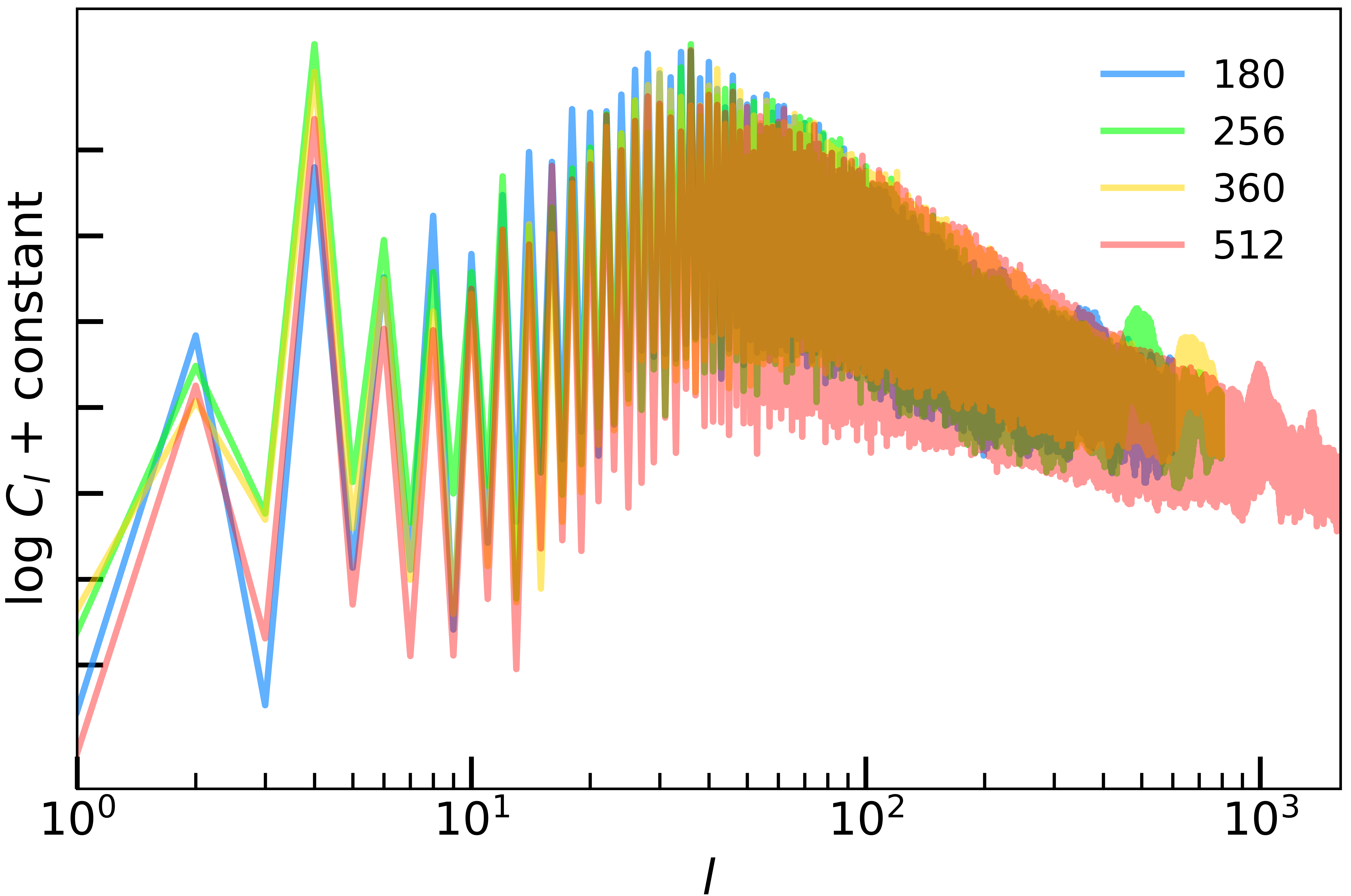}
\caption{Velocity power spectra at four different computational resolutions. The peaks at $l=4$ are generated by a mild carbuncle instability arising in our solutions. A smaller peak can be seen at $l>100$ in the left panel, caused by the grid scale. It occurs at different locations in the plot for different resolutions.}
\label{fig:convergence_full}
\end{figure}

Here we describe the impact of spatial resolution and the duration of numerical calculations on our results. Fig \ref{fig:convergence_full} shows the velocity power spectrum (without additional smoothening) obtained from one of our fiducial models ($n=9,s=0$) at four different spatial resolutions. It can be seen that spectra for all resolutions agree with each other on their general shape and location of the break wavenumber. Moreover, they all have a distinct feature at $l=4$. As previously mentioned, we identify this feature, also seen in Figure \ref{fig:mollweide} as square-like patterns, with a numerical instability (called the carbuncle instability) that arises due to the shock (spherical here) not perfectly aligning everywhere with the grid geometry (Cartesian in this case). Nevertheless, the spurious mode at $l=4$ introduced by this instability is seen to decrease with resolution. No other similar effect is seen on the power spectrum that decreases with resolution and that can therefore be identified with grid effects or the carbuncle instability. A similar phenomenon was observed by \citep{Fraschetti+2010A&A} for their 3D numerical models of SNRs using a Cartesian grid. They used Adaptive Mesh Refinement (AMR) to resolve the shock structure in the SNR and found a similar agreement for the shock structure between the lowest and highest levels of resolution.

The effect from the grid geometry on the power at different modes is the most prominent at early times, as exhibited by Figure \ref{fig:evolution}. At $t\sim40\mathrm{d}$ (or $t/t_i=40$), the spectrum is mostly flat except for features at low values of $l$ ($l\lesssim10$), where it peaks. These features are indicative of grid effects. Even though initially these features have power roughly a couple orders of magnitude greater than power at any other scale, turbulent activity quickly introduces power at all modes (with the largest growth for $l=l_0$). The power spectrum is seen to reach its steady state shape after nearly three decades of time elapsed, and saturation of the power spectrum is attained by four decades. This is in agreement with \cite{Warren+2013MNRAS}, who find that running calculations for at least four decades of time is required for the effects of the initial perturbations to get washed out by saturation of the instabilities.

\bibliographystyle{apj} 
\typeout{}
\bibliography{smbib}

\end{document}